\documentclass[journal]{IEEEtran}
\usepackage{amssymb}
\usepackage{amsmath}
\usepackage[caption=true]{caption}
\usepackage{subcaption}
\usepackage{epsfig}
\usepackage{amsthm}
\usepackage{mathtools}
\usepackage{multirow}
\usepackage{float}
\usepackage{color}
\usepackage{cite}
\usepackage{graphicx}
\usepackage{graphics}
\usepackage{epstopdf}
\usepackage{etoolbox}

\floatstyle{plain}
\newfloat{bottomfigure}{bp}{lop}
\newtheorem{theorem}{Theorem}
\newtheorem{corollary}{Corollary}
\newtheorem{lemma}{Lemma}

\newtheorem{assumption}{Assumption}
\newtheorem{remark}{Remark}

\newtheorem{proposition}{Proposition}

\def\b0{\text{{\bf 0}}}

\begin{document}

\title{Synchronization in Networks of Identical Systems via Pinning:
Application to Distributed Secondary Control of Microgrids}
\author{S. Manaffam$^{1}$, M. K. Talebi$^{2}$, A. K. Jain$^{3}$, and A. Behal%
$^{4}$\thanks{$^{1}$S. Manaffam is with the NanoScience Technology Center,
University of Central Florida (UCF), Orlando, FL 32826. Email: \texttt%
{\small saeedmanaffam@knights.ucf.edu}}\thanks{$^{2}$M. K. Talebi is a
graduate student in the ECE Dept. at University of Central Florida (UCF),
Orlando, FL 32816 }\thanks{$^{3}$A. K. Jain is with the Client Computing
Group at Intel Corporation. in Portland, OR.}\thanks{$^{4}$Aman Behal is
with the ECE\ Dept. and the NanoScience Technology Center, UCF, Orlando, FL
32826. Email: \texttt{\small abehal@ucf.edu} (corresponding author)}}
\maketitle

\begin{abstract}
Motivated by the need for fast synchronized operation of power microgrids,
we analyze the problem of single and multiple pinning in networked systems.
We derive lower and upper bounds on the algebraic connectivity of the
network with respect to the reference signal. These bounds are utilized to
devise a suboptimal algorithm with polynomial complexity to find a suitable
set of nodes to pin the network effectively and efficiently. The results are
applied to secondary voltage pinning control design for a microgrid in
islanded operation mode. Comparisons with existing single and multiple
pinning strategies clearly demonstrate the efficacy of the obtained results.

%\begin{IEEEkeywords}
%	 Pinning control, Secondary control, Distributed complex networks, Networked control systems (NCS), Microgrid, Synchronization.
%	 \end{IEEEkeywords}
\end{abstract}

\section{Introduction}

%The problem of synchronization in a complex network of oscillators has found
%its place in many fields of research such as biology, physics, and
%engineering \cite{Newman10}.
The control of complex networks is a problem that arises quite frequently in
industrial networked systems. One example of these industrial applications
with potential for huge economical impact is smart grids and one of its
integral parts, the microgrid. The U.S. Department of Energy (DoE)
identifies a microgrid as a group of interconnected loads and distributed
generators (DGs) with clearly identifiable electrical boundaries that can be
controlled as a single entity with respect to the main grid, which can connect
and disconnect from the grid as needed, \textit{i. e.}, it can operate in
either grid connected or so-called islanded mode \cite{Khodaei15}\cite%
{Microgrid11}. During islanding, so-called secondary control is needed to
correct the deviation of DG output voltages and grid frequency from their
nominal values. Distributed control with pinning is a practical option to
get the voltage and frequency across the network to synchronize to reference
values that are injected at one or more nodes in the network. If the
reference signal is injected through proper nodes, the inherent connectivity
of the underlying network will force the rest of the nodes to converge to
the reference state. This scheme of networked control system design is
called pinning control and is a very effective method for controlling
distributed systems \cite{Chen07,Sorrentino07,Li04,Yu09,DeLilles11a}.

In \cite{Chen07}, it has been shown under the assumption of positive
definite coupling that a network of oscillators can be stabilized by a
single controller. Clearly, pinning a network in this manner requires a very
large controller gain, which might be practically undesirable if not
impossible. To sidestep this issue, research has evolved to use multiple
controllers with smaller gains to obtain practical design of controllers
utilizing lower number of controllers \cite{Li04,Sorrentino07}. In \cite%
{Li04}, pinning of higher degree nodes has been investigated. In \cite{Yu09}%
, it is proposed to pin the lower degree nodes to stabilize the network
globally. It has been shown that in certain cases, this approach outperforms
that of \cite{Li04}. Adaptive pinning control is used in \cite{Chen07}\cite%
{Yu09}\cite{DeLilles11a}\cite{DeLellis13}, where the controller gains are
chosen to be adaptively governed by differential equations. As the minimum
eigenvalue of the pinned matrix plays an important role in the stability of
the network, there have been some efforts to bound this measure. Recently in 
\cite{Bapat16}, for a weighted tree, it has been shown that the minimum
eigenvalue of the pinned matrix is upper bounded by the algebraic
connectivity\footnote{%
defined as the smallest nonzero eigenvalue of the Laplacian matrix \cite%
{Mohar91}.} of the unpinned network; however, no lower bound is provided. It
should be noted that in most applications, the lower bound of the pinned
Laplacian provides a sufficient condition on synchronization of the network
and the upper bound only provides a necessary condition. In \cite{Pirani16},
both lower and upper bounds for the minimum eigenvalue of the pinned
Laplacian are provided; however, the lower bound provided in \cite{Pirani16}
requires knowledge of the nonnegative eigenvector corresponding to the
minimum eigenvalue of the pinned Laplacian.

In this paper, motivated by a desire to obtain fast voltage synchronization
in a microgrid, we tackle the problem of pinning a network of identical
systems to a given reference signal. First, we derive tight upper and lower
bounds on the algebraic connectivity of the network to the reference signal.
These bounds show that pinning the nodes based on their degrees can be
improved by introducing spectral\footnote{%
The spectrum of a network is the set of eigenvalues of the
Laplacian/gradient matrix of the network \cite{Mohar91}.} measures such as
average path length in conjunction with relative degrees to the pinning set.
According to these findings, we devise a simple suboptimal algorithm with
polynomial complexity to identify the pinning nodes. Finally, we take these
findings and apply them to the problem of voltage control in a microgrid to
achieve fast synchronization to the reference voltage when the microgrid
connection to the main grid is severed. The novelty of this work lies in: 
\textit{(a)} finding close upper and lower bounds on the algebraic
connectivity of pinned network where a single node or multiple nodes are
pinned, \textit{(b)} devising a suboptimal algorithm to localize the pinning
nodes which is shown to converge in polynomial time, and \textit{(c)}
applying the proposed algorithm in developing effective secondary voltage
control of a microgrid in islanded operation mode.

The remainder of this paper is organized as follows. The system model and
control design are described in Section II. Main results are presented in
Section III, and our proposed pinning algorithm is described in Section IV
followed by an illustrative example in Section V. Section VI concludes the
paper.

\section{Preliminaries, Network Model, and Motivation}

\subsection{Preliminaries}

The set of real $n$-vectors is denoted by $\mathbb{R}^{n}$ and the set of
real $m\times n$ matrices is denoted by $\mathbb{R}^{m\times n}$. We refer
to the set of non-negative real numbers by $\mathbb{R}_{+}$. Matrices and
vectors are denoted by capital and lower-case bold letters, respectively. An 
$n\times n$ identity matrix is shown by $\text{\textbf{I}}_{n}$, and 
\begin{equation*}
\text{\textbf{I}}_{n}^{(m)}\triangleq \text{diag}([\underbrace{1~\cdots ~1}%
_{m}\underbrace{0~\cdots ~0}_{n-m}]).
\end{equation*}%
A vector of all ones of size $n$ is denoted by $\text{\textbf{1}}_{n}$ and
its corresponding matrix form by $\text{\textbf{1}}_{n\times n}$. $\text{%
\textbf{1}}_{n\times n}^{(m)}$ is defined as 
\begin{equation*}
\text{\textbf{1}}_{n\times n}^{(m)}\triangleq \left[ 
\begin{array}{ll}
\text{\textbf{1}}_{m\times m} & 0_{m\times N-m-1} \\ 
0_{N-m-1\times m} & 0_{N-m-1\times N-m-1}%
\end{array}%
\right] .
\end{equation*}%
The symmetric part of a matrix, \text{\textbf{X}}, is denoted by $\text{%
\textbf{X}}^{(s)}$ \cite{Manaffam13a}, while $\lambda _{\min }(\cdot )$ and $%
\lambda _{\max }(\cdot )$ denote the minimum and maximum eigenvalue of the
argument square matrix, respectively. The set of vertices/nodes in the
graph/network is denoted by $\text{${\mathcal{N}}$}=\{1,\,\cdots ,\,N\}$.
The network is represented by its adjacency matrix $\text{\textbf{A}}%
=[a_{ij}]$: $a_{ij}=0$ indicates that there is no coupling from node $i$ to
node $j$ and $a_{ij}\neq 0$ indicates a connection from node $j$ to node $i$
with the weight $a_{ij}$. The degree of each node is denoted by $%
d_{i}=\sum_{m=1}^{N}a_{im}(t)$; $d_{\min }$ and $d_{\max }$ refer to minimum
and maximum degrees of the network, respectively. Finally, if the path
length between the nodes $i$ and $j$ is denoted as $\ell (i,j)$, then the
path length between node $i$ and set $\mathcal{J}$ is denoted as $\ell (i,\,%
\mathcal{J})$ 
\begin{equation*}
\ell (i,\,\mathcal{J})\triangleq \min_{j\in \mathcal{J}}\ell (i,j).
\end{equation*}%
We state two preliminary lemmas, which will be used in the reminder of the
paper.

\begin{lemma}
\label{lemma: eigenvalue_bounds} Let $\text{\textbf{A}} \in \mathbb{R}^{n
\times n}$ be a normal matrix, i.e. $\text{\textbf{A}}^T \text{\textbf{A}} = 
\text{\textbf{A}} \text{\textbf{A}}^T$, with eigenvalues $\mu_1\,\ge \,
\mu_2 \, \ge \, \cdots\, \ge\, \mu_N$, then 
\begin{equation}
\mu_1 \text{\textbf{v}}^T \text{\textbf{v}}\, \ge \, \text{\textbf{v}}^T%
\text{\textbf{A}}\text{\textbf{v}} \, \ge \, \mu_N \text{\textbf{v}}^T \text{%
\textbf{v}},
\end{equation}
and if $\text{\textbf{A}}$ is nonsingular, then 
\begin{equation}
\frac{1}{\mu_1} \text{\textbf{v}}^T \text{\textbf{v}}\, \le \, \text{\textbf{%
v}}^T\text{\textbf{A}}^{-1}\text{\textbf{v}} \, \le \, \frac{1}{\mu_N} \text{%
\textbf{v}}^T \text{\textbf{v}}.
\end{equation}
\end{lemma}

%\begin{remark}
%All symmetric matrices are normal. Moreover, the eigenvalues of a normal
%matrix are real, hence they can be sorted as stated in Lemma \ref{lemma:
%eigenvalue_bounds}.
%\end{remark}

\begin{lemma}[Schur complement]
\label{lemma: Schur} The symmetric block matrix 
\begin{eqnarray}
\left[ 
\begin{array}{ll}
\text{\textbf{A}} & \text{\textbf{B}} \\ 
\text{\textbf{B}}^T & \text{\textbf{C}}%
\end{array}
\right]  \notag
\end{eqnarray}
is positive semidefinite if and only if 
\begin{eqnarray}
&(1)& \quad \text{\textbf{A}} \succ \b0,  \notag \\
&(2)& \text{\textbf{C}} - \text{\textbf{B}}^T \text{\textbf{A}}^{-1} \text{%
\textbf{B}} \succeq \b0.  \notag
\end{eqnarray}
\end{lemma}

\subsection{Network Model}

Consider the problem of regulating a network of $N$ linearly coupled
identical systems described by 
\begin{align}
& \text{$\dot{\text{\textbf{x}}}$}_{i}=\text{\textbf{f}}(\text{\textbf{x}}%
_{i})+c\sum_{j=1}^{N}a_{ij}\,\text{\textbf{H}}\,(\text{\textbf{x}}_{j}-\text{%
\textbf{x}}_{i})+\text{\textbf{u}}_{i},  \label{SysModel} \\
& \text{\textbf{x}}_{i}(t_{0})=\text{\textbf{x}}_{i0},  \notag
\end{align}%
where $\text{\textbf{x}}_{i}\in \mathbb{R}^{n}$ is the state vector, $\text{%
\textbf{f}}:\mathbb{R}^{n}\rightarrow \mathbb{R}^{n}$ is a nonlinear
function describing the dynamics of the systems, $\text{\textbf{u}}_i\in 
\mathbb{R}^n$ is the input vector for node $i$, $\text{\textbf{H}}\in 
\mathbb{R}^{n\times n}$ denotes the inner coupling matrix between the states
of coupled nodes, and $c>0$ is the coupling strength. As previously defined, 
$\text{\textbf{A}} = [a_{ij}]$ refers to the adjacency matrix of the network.

The dynamics in \eqref{SysModel} can be rewritten as follows%
\begin{align}
& \text{$\dot{\text{\textbf{x}}}$}_{i}=\text{\textbf{f}}(\text{\textbf{x}}%
_{i})-c\sum_{j=1}^{N}l_{ij}\,\text{\textbf{H}}\,\text{\textbf{x}}_{j}+\text{%
\textbf{u}}_{i},  \label{NetModel} \\
& \text{\textbf{x}}_{i}(t_{0})=\text{\textbf{x}}_{i0},  \notag
\end{align}%
where $\text{\textbf{L}}= [l_{ij}]$~is the Laplacian matrix of the network
and is defined in \cite{Mohar91}. Throughout the paper, we will make the
following assumptions.

\begin{assumption}
\label{assumption: connectivity}The network is undirected, \textit{i.e.,} $%
a_{ij}=a_{ji},$ and connected, \textit{i.e.,} each node can be reached from
any other node in the network.
\end{assumption}

\begin{assumption}
\label{assumption: QUAD}There exists a positive semidefinite matrix \text{%
\textbf{F}}~such that $\Omega \subseteq \mathbb{R}^{n}$ 
\begin{equation}
(\text{\textbf{x}}-\text{\textbf{y}})^{T}\left[ \text{\textbf{f}}(\text{%
\textbf{x}})-\text{\textbf{f}}(\text{\textbf{y}})\right] \leq (\text{\textbf{%
x}}-\text{\textbf{y}})^{T}\text{\textbf{F}}(\text{\textbf{x}}-\text{\textbf{y%
}}),\quad \forall \,\text{\textbf{x}},\text{\textbf{y}}\in \Omega .
\label{eq: boundF}
\end{equation}
\end{assumption}

\begin{remark}
The connectivity of the network implies that $\text{\textbf{L}}$ has a zero
eigenvalue with multiplicity one, which is the result of the Laplacian
matrix having zero-sum rows \cite{Mohar91}. In the case of an undirected
(bidirectional) network, \text{\textbf{L}}~is symmetric and positive
semidefinite \cite{Mohar91}.
\end{remark}

\begin{remark}
Note that Assumption \ref{assumption: QUAD} is not very restrictive, \textit{%
i.e.,} if all elements of the Jacobian of $\text{\textbf{f}}(\cdot )$ on $%
\text{\textbf{x}}\in \Omega $ ~are bounded, then there always exists a
positive semidefinite matrix $\text{\textbf{F}}$ such that Assumption \ref%
{assumption: QUAD} holds \cite{Yu09}. This condition is closely related to
the QUAD condition as discussed in \cite{DeLilles11}. Unlike the QUAD
condition, here, \text{\textbf{F}}~is not necessarily diagonal. Moreover,
the class of functions which satisfies (\ref{eq: boundF}), contains the
class of locally Lipschitz functions \cite{DeLilles11}.
\end{remark}

In order to regulate the network's behavior to converge to the reference
trajectory, a pinning method is used. In the pinning method, nodes are
partitioned into two subsets: (i) nodes without explicit knowledge of the
reference trajectory, and (ii) nodes with direct knowledge of the reference
trajectory called the pinning set. In pinning control, the objective is to
choose the pinning set in a manner that the network converges to the
reference trajectory or state \cite{Sorrentino07}. We assume that the
control input, $\text{\textbf{u}}_{i}$, is chosen as a linear feedback 
\begin{equation}
\text{\textbf{u}}_{i}^{(p)}=-\zeta _{i}g_{i}\text{\textbf{H}}\,\left( \text{%
\textbf{x}}_{i}-\text{\textbf{s}}\right) ,  \label{eq: Controller}
\end{equation}%
where \textbf{s }$\in \mathbb{R}^{n}$ is a reference signal, $g_{i}$ is
pinning gain, and $\zeta _{i}$ is a binary variable indicating if a node is
pinned. Let us define the set of pinning nodes, \text{${\mathcal{P}}$}, as 
\begin{equation*}
\text{${\mathcal{P}}$}\triangleq \{i|~\zeta _{i}=1,~i=1,\cdots ,N\}
\end{equation*}%
and the set of unpinned nodes as $\mathcal{I}=\mathcal{N}\setminus \mathcal{P%
}$. Also, let the pinning matrix, $\text{\textbf{Z}}$ be 
\begin{equation}
\text{\textbf{Z}}\triangleq \text{diag}([\zeta _{1},\cdots ,\zeta _{N}]^{T}).
\label{Pinning Matrix}
\end{equation}%
By defining the tracking error of node $i$ from the reference trajectory as 
%%%%%%%%%%%% 			DEFINITION 2
\begin{equation*}
\text{\textbf{e}}_{i}\triangleq \text{\textbf{x}}_{i}-\text{\textbf{s}},
\end{equation*}%
and assuming $\text{\textbf{u}}_{i}=\text{\textbf{u}}_{i}^{(p)}$, the
closed-loop network dynamics in \eqref{NetModel} can be stated as 
\begin{align}
& \text{$\dot{\text{\textbf{e}}}$}_{i}=\text{\textbf{f}}(\text{\textbf{x}}%
_{i})-\text{\textbf{f}}(\text{\textbf{s}})-\sum_{j=1}^{N}l_{ij}\text{\textbf{%
H}}\text{\textbf{e}}_{j}-\zeta _{i}g_{i}\text{\textbf{H}}\text{\textbf{e}}%
_{i},  \label{eq: NetError} \\
& \text{\textbf{e}}_{i}(t_{0})=\text{\textbf{x}}_{i0}-\text{\textbf{s}}_{0}.
\notag
\end{align}%
Please note that in this formulation, \text{\textbf{Z}}~captures the
locations/nodes in which the reference signal is injected into the network,
while \text{\textbf{G}}~indicates how strongly at each node the reference
signal is injected.

%%%%%%%%%%%%%%%%%%%
%%%%%%%%%%%%%%%%%%%

\subsection{Network Stability}

%%%%%%%%%%%%%%%%%%%

Let us define the overall network tracking error as follows 
\begin{equation*}
\text{\textbf{e}}\triangleq \lbrack \text{\textbf{e}}_{1}^{T}\,\cdots \,%
\text{\textbf{e}}_{N}^{T}]^{T},
\end{equation*}%
and the pinning gain matrix, $\text{\textbf{G}}$, as 
\begin{equation*}
\text{\textbf{G}}=\text{diag}([g_{1},\,\cdots ,\,g_{N}]).
\end{equation*}

\begin{theorem}[Network Stability]
\label{theorem: stability}The network in \eqref{eq: NetError} is
asymptotically stable to $\text{\textbf{e}}=0$, if Assumptions \ref%
{assumption: connectivity} and \ref{assumption: QUAD} hold, and 
\begin{equation}
\text{\textbf{F}}-\mu _{i}\text{\textbf{H}}^{(s)}\prec 0,\quad \forall i\in 
\text{${\mathcal{N}}$}  \label{eq: stab_cond}
\end{equation}%
where $\mu _{i}$ $\forall i\in {\mathcal{N}}$ are eigenvalues of $c\,\text{%
\textbf{L}}+\text{\textbf{Z}}\,\text{\textbf{G }}$\cite{Manaffam13b}.
\end{theorem}

One implication of Theorem \ref{theorem: stability} is as follows:

\begin{proposition}
Let \text{\textbf{H}}~be positive definite, then for a desired convergence
rate of the network, $\alpha$, there exists a pair $(\text{\textbf{Z}}, 
\text{\textbf{G}})$ such that the control law \eqref{eq: Controller}
guarantees an equal or greater rate.
\end{proposition}

The existence of such a pair is evident by setting $\text{\textbf{Z}}=\text{%
\textbf{I}}_{N}$ and $\text{\textbf{G}}=\Big(\left( \alpha +\lambda _{\max }(%
\text{\textbf{F}})\right) /\lambda _{\min }(\text{\textbf{H}})\Big)\,\text{%
\textbf{I}}_{N}$.

\subsection{Problem Formulation}

\begin{figure*}[t]
{\normalsize \ 
\begin{equation}  \label{eq: Laplacian_permutation}
\text{\textbf{L}} = \left[ 
\begin{array}{ccccccc}
\text{\textbf{L}}_0 + \text{\textbf{D}}_0 & - \text{\textbf{B}}_0 & \b0 & \b0
& \b0 & \cdots & \b0 \\ 
- \text{\textbf{B}}_0^T & \text{\textbf{D}}^\prime_0 + \text{\textbf{L}}_1 + 
\text{\textbf{D}}_1 & - \text{\textbf{B}}_1 & \b0 & \b0 & \cdots & \b0 \\ 
\b0 & - \text{\textbf{B}}_1^T & \text{\textbf{D}}_1^\prime+ \text{\textbf{L}}
_2 + \text{\textbf{D}}_2 & - \text{\textbf{B}}_2 & \b0 & \cdots & \b0 \\ 
\b0 & \b0 & -\text{\textbf{B}}_2^T & \text{\textbf{D}}_2^\prime + \text{ 
\textbf{L}}_3 + \text{\textbf{D}}_3 & \text{\textbf{B}}_3 & \ddots & \vdots
\\ 
\vdots & \vdots & \ddots & \ddots & \ddots & \ddots & \b0 \\ 
\b0 & \b0 & \ddots & \ddots & -\text{\textbf{B}}_{k-1}^T & \text{\textbf{D}}
^\prime_{k-1} + \text{\textbf{L}}_{k-1} + \text{\textbf{D}}_k & -\text{ 
\textbf{B}}_{k} \\ 
\b0 & \b0 & \b0 & \cdots & \b0 & -\text{\textbf{B}}_{k}^T & \text{\textbf{L}}
_k + \text{\textbf{D}}_k^{\prime}%
\end{array}
\right]
\end{equation}
\begin{eqnarray}
\mu_u = \frac12 \left( m + g + \frac{m}{N - 1} \right) \left(1 - \sqrt{1 - 
\frac{4 m g}{( N - 1) \left( m + g + \frac{m}{N - 1} \right)^2}} \right).
\label{eq: upper_single}
\end{eqnarray}
\begin{eqnarray}  \label{eq: upper_multiple}
\mu_u = \frac{(N - m_0) (g + d_{ 0, \, \max} + m_0) + \sum\limits_{i =
1}^{m_1} d_{0, \, i}^\prime}{2 (N - m_0)} - \sqrt{\left( \frac{(N-m_0) (g +
d_{ 0, \, \max} + m_0)-\sum\limits_{i = 1}^{m_1} d_{0, \, i}^\prime }{2 (N -
m_0)}\right)^2 + \frac{\sum\limits_{i = 1}^{m_0} d_{0,\, i}^2 }{N-m_0}}
\end{eqnarray}
\hrulefill }
\end{figure*}
As is evident from Theorem \ref{theorem: stability}, the minimum eigenvalue
of $c\,\text{\textbf{L}}+\text{\textbf{Z}}\,\text{\textbf{G}}$ plays an
important role in the stability and the rate of convergence of the network.
Hence, identifying the best pinning strategy is an inseparable part of
controlling and regulating networked systems. In this section, we formulate
two related problems, namely, finding the optimal locations to pin a
specified number of nodes, and identifying the minimum number of nodes
needing to be pinned to guarantee a certain convergence rate to desired
reference trajectory or state:

\begin{enumerate}
\item {identifying the optimal location for pinning nodes: let $\text{%
\textbf{G}}=g\text{\textbf{I}}_{N}$ and the number of desired pinning nodes
be $m$, then find $\text{\textbf{Z}} = \text{diag}([\zeta _{1}\cdots \zeta
_{N}])~~\zeta _{i}\in\{0,\,1\}~\forall i\in \mathcal{N}$ such that $\phi(%
\text{\textbf{Z}}) \triangleq \lambda_{\min}(\text{\textbf{I}}_{N}\otimes 
\text{\textbf{F}}-(c\text{\textbf{L}}+ \text{\textbf{Z}}\text{\textbf{G}}%
)\otimes \text{\textbf{H}})$ is maximized: 
\begin{equation}
\begin{array}{lll}
\text{\textbf{Z}}^\star = & \text{argmax} & \phi(\text{\textbf{Z}}) \\ 
~ & \text{s. t.} & \|\text{\textbf{Z}}\|_0 = m%
\end{array}
\label{eq: Problem1}
\end{equation}
where $\|\cdot\|_0$ denotes norm $0$. }

\item {pinning the minimum number of nodes to achieve a certain convergence
rate, $\lambda^\star$ 
\begin{equation}
\begin{array}{ll}
\min & \|\text{\textbf{Z}}\|_0 \\ 
\text{s. t.} & \phi(\text{\textbf{Z}}) \ge \lambda^\star%
\end{array}
\label{eq: Problem2}
\end{equation}
}
\end{enumerate}

As is well known, these problems are NP-hard \cite{DeLellis13}. In the next
section, we will first calculate tight bounds on the eigenvalues of $c\text{%
\textbf{L}}+\text{\textbf{Z}}\text{\textbf{G}}$. Then, based on the acquired
bounds, we will produce suboptimal algorithms to solve the problems in %
\eqref{eq: Problem1} and \eqref{eq: Problem2} in polynomial time.

\section{Main Results}

We begin by first analyzing the problem of single pinning in depth and
derive its limitations on stability and convergence rate of the network.
Next, the generalization of our analysis for the case of multiple pinning
will be given. Without loss of generality, in the rest of the paper, the
coupling coefficient is assumed to be $c=1$.

\subsection{Single pinning}

In single pinning, the reference trajectory is assumed to be available only
in one of the nodes. For the convenience of analysis and without loss of
generality, we assume that the Laplacian matrix of the network is permuted
as \eqref{eq: Laplacian_permutation}, where $\text{\textbf{L}}_{0}=0$, $%
\text{\textbf{D}}_{0}=m$ and $\text{\textbf{B}}_{0}=\text{\textbf{1}}%
_{m}^{T} $. The next theorem provides suitable upper and lower bounds for
the case of single pinning.

\begin{theorem}
\label{theorem: bounds_single_pinning} Let $m$ be the degree of the pinned
node with pinning gain $g$ in a network of size $N$. If \text{\textbf{L}}~is
the Laplacian matrix of a connected undirected network permuted as 
\eqref{eq:
Laplacian_permutation}, then the maximum $\mu $ such that $\text{\textbf{L}}%
+g\text{\textbf{I}}_{N}^{(1)}-\mu \text{\textbf{I}}_{N}\succeq 0$ belongs to
the interval $[\mu _{l},\,\mu _{u}]$, where the upper bound $\mu _{u}$ is
given in \eqref{eq: upper_single} and the lower bound $\mu _{l}$ is the
smallest positive root of the polynomials, $\alpha _{i}(\mu )$ $i=0,\cdots
,\,k-1$%
\begin{equation}
\alpha _{i}(\mu )=d_{i-1,\min }^{\prime }+d_{i,\,\min }-\mu -{d^{\prime }}%
_{i,\max }{d}_{i,\max }/\alpha _{i+1}(\mu ),  \label{eq: lower_single}
\end{equation}%
where 
\begin{eqnarray*}
\alpha _{k}(\mu ) &=&d_{k,\,\min }^{\prime }-\mu \\
d_{i,\,\min } &=&\min (\text{\textbf{B}}_{i}\text{\textbf{1}}) \\
d_{i,\,\min }^{\prime } &=&\min (\text{\textbf{B}}_{i}^{T}\text{\textbf{1}})
\\
d_{i,\,\max }^{\prime } &=&\max (\text{\textbf{B}}_{i}^{T}\text{\textbf{1}})
\\
d_{-1,\,\min }^{\prime } &=&g,\quad d_{0,\,\min }=d_{0,\,\max }=m.
\end{eqnarray*}%
and $k$ denotes the path-length of the farthest node to the pinning node.%
%\begin{figure*}[t]
%{\normalsize \ 
%
%}
%\par
%{\normalsize \hrulefill \vspace*{4pt} }
%\end{figure*}

\begin{proof}
\textbf{Part A (upper bound): }The Laplacian matrix in 
\eqref{eq:
Laplacian_permutation} can be written as 
\begin{equation*}
\text{\textbf{L}}=\left[ 
\begin{array}{ll}
m & 
\begin{array}{ll}
-\text{\textbf{1}}_{m}^{T} & 0_{N-m-1}^{T}%
\end{array}
\\ 
\begin{array}{l}
-\text{\textbf{1}}_{m} \\ 
0_{N-m-1}%
\end{array}
& \text{\textbf{L}}_{1}+\text{\textbf{I}}_{N-1}^{(m)}%
\end{array}%
\right] ,
\end{equation*}%
where $\text{\textbf{L}}_{1}$ is an $N-1\times N-1$ Laplacian matrix. Using
Lemma \ref{lemma: Schur}, $\text{\textbf{L}}+g\text{\textbf{I}}%
_{N}^{(1)}-\mu \text{\textbf{I}}_{N}\succeq 0$ iff 
\begin{eqnarray*}
m+g-\mu &>&0, \\
\text{\textbf{W}}\triangleq \text{\textbf{L}}_{1}+\text{\textbf{I}}%
_{N-1}^{(m)}-\mu \,\text{\textbf{I}}_{N-1}-\frac{1}{m+g-\mu }\text{\textbf{1}%
}_{N-1\times N-1}^{(m)} &\succeq &0.
\end{eqnarray*}%
Using Lemma \ref{lemma: eigenvalue_bounds} with $\text{\textbf{v}}=\text{%
\textbf{1}}_{N-1}$, we have 
\begin{eqnarray}
m+g-\mu &>&0 \\
m-(N-1)\mu -\frac{m^{2}}{m+g-\mu }\geq \lambda _{\min }(\text{\textbf{W}})
&\geq &0
\end{eqnarray}%
Solving the resultant quadratic equation and taking into account that $%
m+g>\mu $, the upper bound in \eqref{eq: upper_single} can be obtained.

\textbf{Part B (lower bound):} Without loss of generality, we assume that
the Laplacian matrix of the network can be permuted to 
\eqref{eq:
Laplacian_permutation} where $d_{i,\min }$ and $d_{i,\,\min }^{\prime }$ are
minimum of the nonzero entries of $\text{\textbf{D}}_{i}\triangleq \text{diag%
}(\text{\textbf{B}}_{i}\text{\textbf{1}})$ and $\text{\textbf{D}}%
_{i}^{\prime }\triangleq \text{diag}(\text{\textbf{B}}_{i}^{T}\text{\textbf{1%
}})\succeq \text{\textbf{I}}$. Define $\text{\textbf{Y}}_{i}$ as 
\begin{equation*}
\text{\textbf{Y}}_{i}=\text{\textbf{L}}_{i}+\text{\textbf{D}}_{i-1}^{\prime
}+\text{\textbf{D}}_{i}-\mu \text{\textbf{I}}-\text{\textbf{B}}_{i+1}\text{%
\textbf{Y}}_{i+1}^{-1}\text{\textbf{B}}_{i+1}^{T},\quad i=0,\,\cdots ,\,k
\end{equation*}%
where $\text{\textbf{B}}_{k+1}=0$, $\text{\textbf{D}}_{k}=0$, $\text{\textbf{%
D}}_{0}^{\prime }=\text{\textbf{I}}_{m}$, $\text{\textbf{D}}_{-1}^{\prime
}=0 $, $\text{\textbf{L}}_{0}=g$, and $\text{\textbf{D}}_{0}=m$. Employing
Lemma \ref{lemma: Schur}, the conditions on $\text{\textbf{L}}+g\text{%
\textbf{I}}_{N}^{(1)}-\mu \text{\textbf{I}}_{N}\succeq 0$ become 
\begin{equation*}
\text{\textbf{Y}}_{i}\succ 0\quad i=0,\cdots ,\,k.
\end{equation*}%
From Weyl's inequalities \cite{Manaffam13a}, the lower bound on the minimum
eigenvalues of the $\text{\textbf{Y}}_{i}$s are 
\begin{equation*}
\lambda _{\min }(\text{\textbf{Y}}_{i})\!\!\geq d_{i-1,\min }^{\prime
}+d_{i,\,\min }-\mu -\lambda _{\max }(\text{\textbf{B}}_{i+1}\text{\textbf{B}%
}_{i+1}^{T})/\lambda _{\min }(\text{\textbf{Y}}_{i+1}),
\end{equation*}%
Since $\lambda _{\max }(\text{\textbf{B}}_{i}\text{\textbf{B}}%
_{i}^{T})=\Vert \text{\textbf{B}}_{i}\Vert _{2}^{2}\leq \Vert \text{\textbf{B%
}}_{i}\Vert _{1}\Vert \text{\textbf{B}}_{i}\Vert _{\infty }={d^{\prime }}%
_{i,\max }{d}_{i,\max }$, 
\begin{align*}
& \lambda _{\min }(\text{\textbf{Y}}_{i}) \\
& \geq d_{i-1,\min }^{\prime }+d_{i,\,\min }-\mu -{d^{\prime }}_{i+1,\max }{d%
}_{i+1,\max }/\lambda _{\min }(\text{\textbf{Y}}_{i+1}).
\end{align*}%
Define $\alpha _{i}(\mu )=d_{i-1,\min }^{\prime }+d_{i,\,\min }-\mu -{%
d^{\prime }}_{i+1,\max }{d}_{i+1,\max }/\alpha _{i+1}(\mu )$, with $\alpha
_{k}(\mu )=d_{k,\,\min }^{\prime }-\mu $.
\end{proof}
\end{theorem}

\begin{corollary}
The connectivity of the network with respect to the reference signal with
single pinning is always less than $d^{\max }/(N-1)$,\textit{\ i.e.,} $\mu
_{u}<d^{\max }/(N-1)\leq 1$. Furthermore, if the pinning gain satisfies, $%
g\gg d^{\max }$ or $g\ll d^{\max }$, then 
\begin{equation*}
\mu _{u}\approx \frac{mg}{Nm+(N-1)g}<\frac{d^{\max }}{N-1}.
\end{equation*}
\end{corollary}

\subsection{Multiple Pinning}

In this section, we assume that $m_{0}$ number of nodes are pinned. Similar
to the case of single pinning, let the Laplacian matrix be permuted as 
\eqref{eq:
Laplacian_permutation}, where $d_{i,\min }$ and $d_{i,\,\min }^{\prime }$
are minimum nonzero entries of $\text{\textbf{D}}_{i}\triangleq \text{diag}(%
\text{\textbf{B}}_{i}\text{\textbf{1}})$ and $\text{\textbf{D}}_{i}^{\prime
}\triangleq \text{diag}(\text{\textbf{B}}_{i}^{T}\text{\textbf{1}})\succeq 
\text{\textbf{I}}_{m_{i+1}}$, $m_{i}$ is the number of rows in the $i^{th}$
block, and $k$ is the path-length of the farthest node to the pinning set: $%
m_{0}+\sum\limits_{i=1}^{k}m_{i}=N$. $d_{i,\,j}$ and $d_{i,\, j}^\prime$
refer to j$th$ diagonal entry of $\text{\textbf{D}}_i$ and $\text{\textbf{D}}%
_i^\prime$, respectively.

\begin{theorem}
\label{theorem: bounds_multiple_pinning} Let \text{\textbf{L}}~ be connected
and permuted as \eqref{eq: Laplacian_permutation}. If $m_{0}$ number of
nodes are pinned, then the maximum $\mu $ such that $\text{\textbf{L}}+g%
\text{\textbf{I}}_{N}^{(m_{0})}-\mu \text{\textbf{I}}_{N}\succeq 0$ belongs
to the interval $[\mu _{l},\,\mu _{u}]$, the upper bound $\mu _{u}$ is given
in \eqref{eq:
upper_multiple}, and the lower bound $\mu _{l}$ is the smallest positive
root of the polynomials, $\alpha _{i}(\mu )$ $i=0,\cdots ,\,k-1$ 
\begin{equation}
\alpha _{i}(\mu )=d_{i-1,\min }^{\prime }+d_{i,\,\min }-\mu -{d^{\prime }}%
_{i,\max }{d}_{i,\max }/\alpha _{i+1}(\mu ),  \label{eq: lower_multiple}
\end{equation}%
where 
\begin{eqnarray*}
\alpha _{k}(\mu ) &=&d_{k,\,\min }^{\prime }-\mu \\
d_{i,\,\min } &=&\min (\text{\textbf{B}}_{i}\text{\textbf{1}}) \\
d_{i,\,\min }^{\prime } &=&\min (\text{\textbf{B}}_{i}^{T}\text{\textbf{1}})
\\
d_{i,\,\max }^{\prime } &=&\max (\text{\textbf{B}}_{i}^{T}\text{\textbf{1}})
\\
d_{-1,\,\min }^{\prime } &=&g.
\end{eqnarray*}%
\begin{IEEEproof}
	Proof is similar to that of Theorem \ref{theorem: bounds_single_pinning}.
	\end{IEEEproof}
\end{theorem}

It should be noted that in \eqref{eq: upper_multiple}, the term $%
\sum\limits_{i=1}^{m_{1}}d_{0,\,i}^{\prime }$ is the number of connections
from the pinning set to the rest of the network and is a measure of the
connectivity of the pinning set to the rest of the network.

\begin{corollary}
\label{corollary: minimum_pinning} The minimum eigenvalue of $\text{\textbf{L%
}}+g\text{\textbf{I}}_{N}^{(m_{0})}$ is upper bounded by 
\begin{equation*}
\lambda _{\min }(\text{\textbf{L}}+g\text{\textbf{I}}_{N}^{(m_{0})})<m_{0}.
\end{equation*}%
\begin{IEEEproof}
	 Setting the upper bounds $\sum\limits_{i = 1}^{m_1} d_{0, \, i}^\prime \le m_0 (N - m_0) $, $d_{0, \, \min} \le N - m_0$, and $\sum\limits_{i = 1}^{m_0} d_{0,\, i}^2 \le m_0 (N - m_0)^2$ in \eqref{eq: upper_multiple}, the proof follows. 
	\end{IEEEproof}
\end{corollary}

\section{Suboptimal Algorithm for Pinning}

\subsection{Algorithm to pin $m$ nodes}

Based on Theorems \ref{theorem: bounds_single_pinning} and \ref{theorem:
bounds_multiple_pinning} as well as the respective corollaries, we propose
to maximize the following objective function to capture the behavior of the
algebraic connectivity, $\mu _{N}$,%
\begin{equation}
f_{i}=\mu _{u}+\mu _{l}-\frac{1}{|\text{${\mathcal{I}}$}\setminus \{i\}|}%
\sum_{{\scriptsize j\in \text{${\mathcal{I}}$}\setminus \{i\}}}\ell (i,\,j).
\label{eq: objective}
\end{equation}%
The first term in the objective function implies that increasing the number
of outgoing connections from the pinning set, $\sum%
\limits_{i=1}^{m_{1}}d_{0,\,i}^{\prime }$, increases the algebraic
connectivity, $\mu _{N}$ (here, $m_{1}$ is the number of immediate neighbors
of the pinning set). The second and third terms imply that minimizing the
distance of the pinning set from the farthest node in the network, $k$,
and/or average path length of the candidate node to the unpinned set, $\text{%
${\mathcal{I}}$}$, increases the lower bound, $\mu _{l}$, which in turn
increases $\mu _{N}$. Thus, our algorithm to find the best $m_{0}$ (1$\leq
m_{0}\leq N$) nodes to pin with respect to the objective function in (\ref%
{eq: objective}) can be explicitly stated as follows

\begin{enumerate}
\item {set: $\text{${\mathcal{P}}$} = \emptyset, \,\text{${\mathcal{I}}$} = 
\text{${\mathcal{N}}$} $, }

\item {while $|\text{${\mathcal{P}}$}|<m_{0}$}

\begin{itemize}
\item {$j=\text{argmax}\{f_{i}|\,\forall i\in \text{${\mathcal{I}}$}\}$}

\item {$\text{${\mathcal{P}}$}=\text{${\mathcal{P}}$}\bigcup \{j\}$, $\text{$%
{\mathcal{I}}$}=\text{${\mathcal{N}}$}\setminus \text{${\mathcal{P}}$}$ }
\end{itemize}
\end{enumerate}

The complexity of calculating the upper and lower bounds in (\ref{eq:
objective}) is $\mathcal{O}(1)$ and $\mathcal{O}(N)$, respectively, whereas
the complexity of computing the third term is $\mathcal{O}(N^{2})$.
Furthermore, the number of searches in the proposed algorithm is $%
m_{0}(N-(m_{0}-1)/2)$ which is a linear function of network size. The total
complexity of the proposed algorithm at its peak $m_{0}=N/2$ is $\mathcal{O}%
(N^{4})$ . On the other hand, the complexity of the search for the optimal
solution can be shown (by Stirling's approximation for large $N$) to scale
exponentially by $N$, 
\begin{equation*}
\left( 
\begin{array}{c}
N \\ 
m_{0}%
\end{array}%
\right) \sim \frac{N^{N}}{m_{0}^{m_{0}}(N-m_{0})^{(N-m_{0})}}.
\end{equation*}%
For $m_{0}=N/2$, the search complexity becomes $2^{N}$; furthermore, each
search involves calculating the minimum eigenvalue. Thus, the optimal
solution to the pinning problem is NP-hard \cite{Pirani16,Manaffam13b}.
Compared to the exponential complexity of the optimal solution, the proposed
algorithm is a great improvement for the slight loss in performance as will
be illustrated in the next section.

\subsection{Algorithm to achieve a desired pinning connectivity, $\protect%
\mu ^\star$}

From Corollary \ref{corollary: minimum_pinning}, we know that the minimum
number of pinning nodes to achieve a targeted algebraic connectivity to
pinning set, $\mu ^{\star }$ is lower bounded by $m_{0}=\lfloor \mu ^{\star
}\rfloor +1$ \footnote{$\lfloor \cdot \rfloor $ denotes the floor operator
and returns the largest previous integer number to the argument.}, hence the
algorithm to find the pinning set to achieve $\mu ^{\star }$ can be devised
as

\begin{enumerate}
\item {$m_0 = \lfloor \mu^\star \rfloor+1 $ }

\item {\ }

\begin{enumerate}
\item {set: $\text{${\mathcal{P}}$}=\emptyset ,\,\text{${\mathcal{I}}$}=%
\text{${\mathcal{N}}$}$, }

\item {while $|\text{${\mathcal{P}}$}|<m$ }

\begin{itemize}
\item {$j = \text{argmax}\{f_i | \, \forall i \in \text{${\mathcal{I}}$}\}$}

\item {$\text{${\mathcal{P}}$} =\text{${\mathcal{P}}$} \bigcup \{ j\}$, $%
\text{${\mathcal{I}}$} = \text{${\mathcal{N}}$} \setminus \text{${\mathcal{P}%
}$}$ }
\end{itemize}
\end{enumerate}

\label{step: choose_node}

\item {if $\text{\textbf{L}} + g \tilde{\text{\textbf{I}}}_N^{{\scriptsize (%
\text{${\mathcal{P}}$})}} - \mu^\star \text{\textbf{I}}_N \succeq \b0$, then
stop;}

\item {set $m_0 = m_0 + 1$ and go to \ref{step: choose_node}.}
\end{enumerate}

The aforementioned bounds and algorithms provide insight into the effect of
choosing the locations and gains to inject the reference signal into the
network of dynamical systems to achieve the desired performance. It can be
seen that the desired criteria of suitable nodes for pinning include large
number of connections as well as smaller maximum distance of the pinning
node(s) from the rest of the network. The results also provide a guideline
for choosing a communication network (if there are no constraints on the
establishing the links), \textit{viz.,} the number of links from the pinning
set to the rest of the network, $\text{\textbf{B}}_{0}$, should be as large
as possible. One advantage of this approach is that it not only provides a
better convergence rate, but it also adds to the robustness of the network
to link failures, \textit{i.e.,} the network is less likely to become
disconnected and/or unstable, if one or more of the links in the
communication network fails (hardware failure, packet drop out, \textit{etc.}%
).

\begin{table*}[t]
\centering
\begin{tabular}{|l|c|c|c|c|c|c|l|}
\hline
\text{Algorithm} & \text{Pinning Node, $i$} & \text{Degree} & $\mu_l$, (\ref%
{eq: lower_single}) & $\mu_N$ & $\mu_u$, (\ref{eq: upper_single}) & $\bar{%
\ell}(i,\mathcal{I})$ & $f_i$ \\ \hline\hline
\text{Optimal, Proposed, and Highest degree} & 14 & 8 & 0 & 0.434 & 0.570 & 
1.540 & -0.970 \\ 
\text{Highest in-betweenness, and Highest centrality} & 2 & 7 & 0 & 0.391 & 
0.503 & 1.538 & -1.035 \\ 
\text{Lowest degree} & 1 & 1 & 0 & 0.065 & 0.076 & 2.462 & -2.385 \\ \hline
\end{tabular}%
%
%\vspace{-.4cm}
\caption{Single pinning, $g=100$, $\protect\mu _{N}=\protect\lambda _{\min }(%
\text{\textbf{L}}+\text{\textbf{Z}}\text{\textbf{G}})$ and $\bar{\ell}(i,%
\mathcal{I})\triangleq \frac{1}{|\mathcal{I}|}\sum\limits_{%
{\protect\scriptsize j\in \mathcal{I}}}\ell (i,\,j)$.}
\label{table: single pinning}
\end{table*}

\section{Case Study: Distributed Control in Microgrid}

\begin{figure}[t]
\centering
\includegraphics[scale=.27]{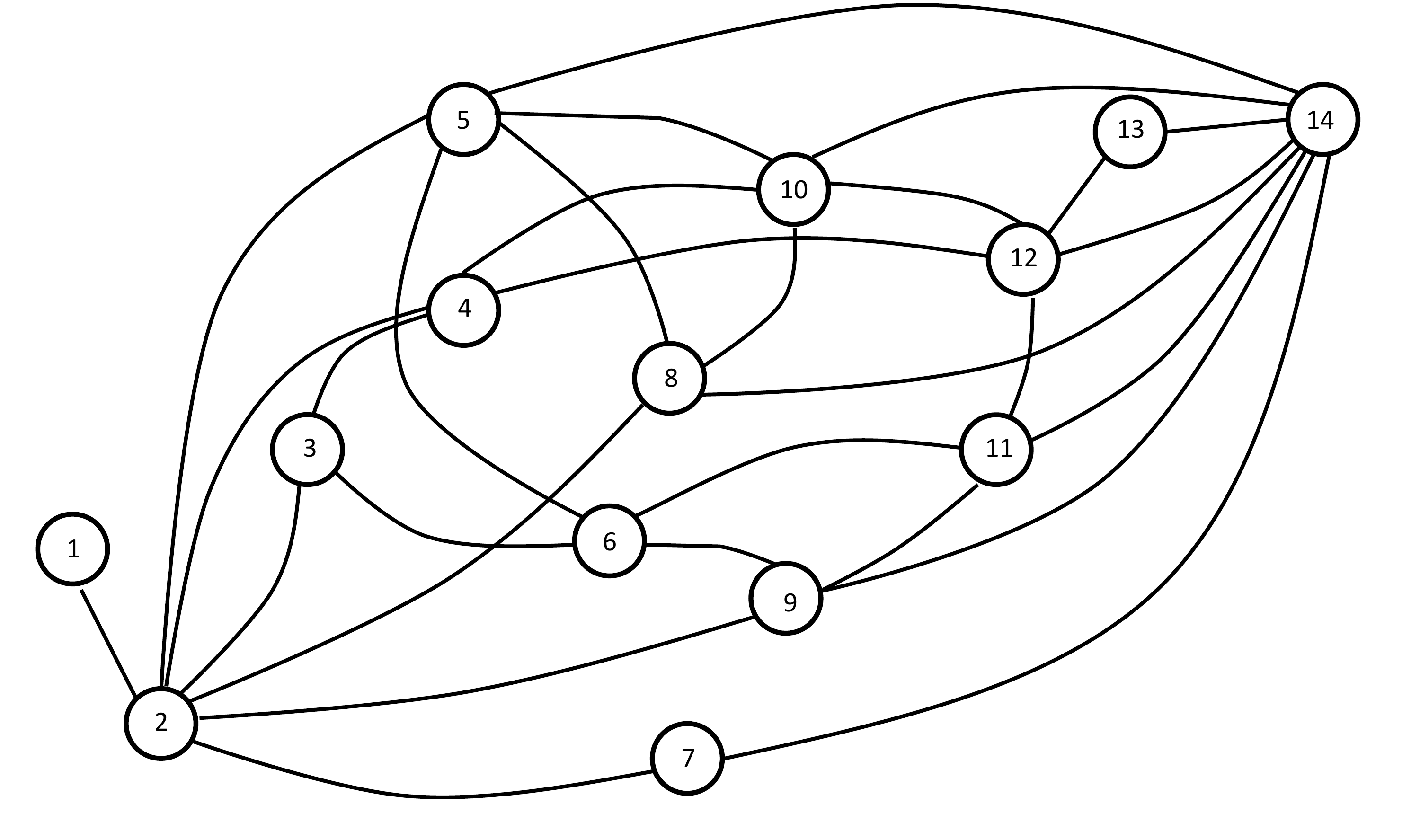} 
\caption{Network topology for DG's with the degree sequence $%
\{1,\,7,\,3,\,4,\,5,\,4,\,2,\,4,\,4,\,5,\,4,\,5,\,2,\, 8\}$.}\vspace{-.4cm}
\label{fig: topology}
\end{figure}

To verify the results in previous sections, we consider an islanded network
of distributed generators (DGs) given in Fig. \ref{fig: topology}. The
dynamics of primary control,\textit{\ i.e.,} droop control, for $i^{th}$ DG
is \cite{Bidram13a} 
\begin{equation}
v_{i}=V_{ni}-n_{Q_{i}}Q_{i},  \label{eq: droop}
\end{equation}%
where $v_{i}$ is the reference value for the output voltage, $V_{ni}$ is the
reference for primary control, $n_{Q_{i}}$ is the droop coefficient. For
further details on the internal dynamics of DGs, please refer to \cite%
{Bidram13a}. Time differentiating the droop equation in \eqref{eq:
droop} obtains %, we have 
\begin{equation*}
\text{$\dot{v}$}_{i}=\dot{V}_{ni}-n_{Q_{i}}\dot{Q}_{i}=u_{i},
\end{equation*}%
where $u_{i}$ is the output of secondary voltage control, $n_{Q_{i}}$ is a
droop coefficient, $\dot{Q}_{i}=-\omega _{c}Q_{i}+Q_{instant,i}$, $\omega
_{c}=12\pi $ and the instantaneous output reactive power, $Q_{instant,i}$,
is measurable. Here, we assume that the secondary voltage control is
distributed and the communication network is the same as the plant network, 
\textit{i.e.,} if DG $i$ and DG $j$ are connected in the grid, then $v_{i}$
and $v_{j}$ are known at both DGs. This is a practical assumption as these
values can be communicated by power line communication (PLC). Thus, the
secondary voltage controller for $i^{th}$ DG can be written as 
\begin{eqnarray}
u_{i} &=&-k\sum_{j\in \text{${\mathcal{N}}$}}a_{ij}(v_{i}-v_{j})-g\zeta
_{i}(v_{i}-v_{{\scriptsize ref}})  \notag \\
&=&-k\left( \sum_{j\in \text{${\mathcal{N}}$}}a_{ij}(v_{i}-v_{j})-g^{\prime
}\zeta _{i}(v_{i}-v_{{\scriptsize \text{ref}}})\right)  \label{eq: secondary}
\end{eqnarray}%
where $k$ is distributed controller gain, $g$ is pinning gain, and $v_{%
{\scriptsize ref}}$ is the reference value for the output voltage. Let the
error from reference be $e_{i}\triangleq v_{i}-v_{{\scriptsize ref}}$; since 
$\dot{v}_{{\scriptsize ref}}=0$, the dynamics of the error can be expressed
as 
\begin{equation}
\text{$\dot{\text{\textbf{e}}}$}=-k(\text{\textbf{L}}+\text{\textbf{Z}}\,%
\text{\textbf{G}}^{\prime })\,\text{\textbf{e}},  \label{eq: DCM}
\end{equation}%
where $\text{\textbf{e}}\triangleq \lbrack e_{1},\,e_{2},\,\cdots
,\,e_{N}]^{T}$. From this equation, we can see that the rate of convergence
for output reference for the primary control is $k\mu $ where $\mu $ is the
algebraic connectivity to auxiliary reference $v_{{\scriptsize ref}}$,%
\textit{\ i.e.,} minimum eigenvalue of $\text{\textbf{L}}+\text{\textbf{Z}}\,%
\text{\textbf{G}}^{\prime }$. In the rest of this section, the distributed
and pinning gains will be assumed to be $k=10$ and $g=100$, respectively.
Furthermore, the desired output voltage is assumed to be $v_{{\scriptsize ref%
}}=380$[Vrms].

\begin{table*}[tbp]
\centering
\begin{tabular}{|l|c|c|c|c|c|l|l|l|}
\hline
\text{Algorithm} & \text{Pinning Set, \text{${\mathcal{P}}$}} & \text{Degrees%
} & $\mu_l$, (\ref{eq: lower_multiple}) & $\mu_N$ & $\mu_u$, (\ref{eq:
upper_multiple}) & $\bar{\ell}(\text{${\mathcal{P}}$},\text{${\mathcal{I}}$}%
) $ & $f_{\mathcal{P}}$ &  \\ \hline\hline
\text{Optimal} & \{ 1, 4, 6, 7, 8, 11, 13\} & 1, 4, 4, 2, 4, 4, 2 & 1.718 & 
2.460 & 2.640 & 1 & 3.35 &  \\ 
\text{Proposed} & \{1, 2, 3, 6, 10, 12, 14\} & 1, 7, 3, 4, 5, 5, 8 & 1.715 & 
1.970 & 2.880 & 1 & 3.60 &  \\ 
\text{Lowest degrees} & \{1, 3, 6, 7, 9, 11, 13\} & 1, 3, 4, 4, 2, 4, 2 & 
0.000 & 1.270 & 1.680 & 1.286 & 0.40 &  \\ 
\text{Highest degrees} & \{2, 5, 9, 10, 11, 12, 14\} & 7, 5, 4, 5, 4, 5, 8 & 
0.721 & 0.990 & 2.220 & 1 & 1.94 &  \\ 
\text{Highest centrality} & \{2, 4, 5, 8, 9, 10, 14\} & 7, 4, 5, 4, 4, 5, 8
& 0.788 & 0.990 & 1.810 & 1 & 1.60 &  \\ 
\text{Highest in-betweenness} & \{2, 4, 5, 6, 9, 12, 14\} & 7, 4, 5, 4, 4,
5, 8 & 0.721 & 0.990 & 2.630 & 1 & 2.35 &  \\ \hline
\end{tabular}%

\caption{Multiple pinning scenario, $g= 100$, $m = 7$, and $\bar{\ell}(%
\mathcal{P},\mathcal{I}) \triangleq \frac1{|\mathcal{I}|}\sum\limits_{ 
{\protect\scriptsize j \in \mathcal{I} }} \ell (j, \, \mathcal{P})$.}
\label{table: multiple pinning}\vspace{-.5cm}
\end{table*}

\subsection{Single Pinning}

\begin{figure}[tbp]
\includegraphics[scale = 0.26]{Fig2}\caption{Evolution of voltage signals of the microgrid with optimal pinning, 
	$\protect\zeta _{14}=1$ at $t=1\,sec$.}
\label{fig: SVT} \vspace{.2cm} %
\includegraphics[scale = 0.29]{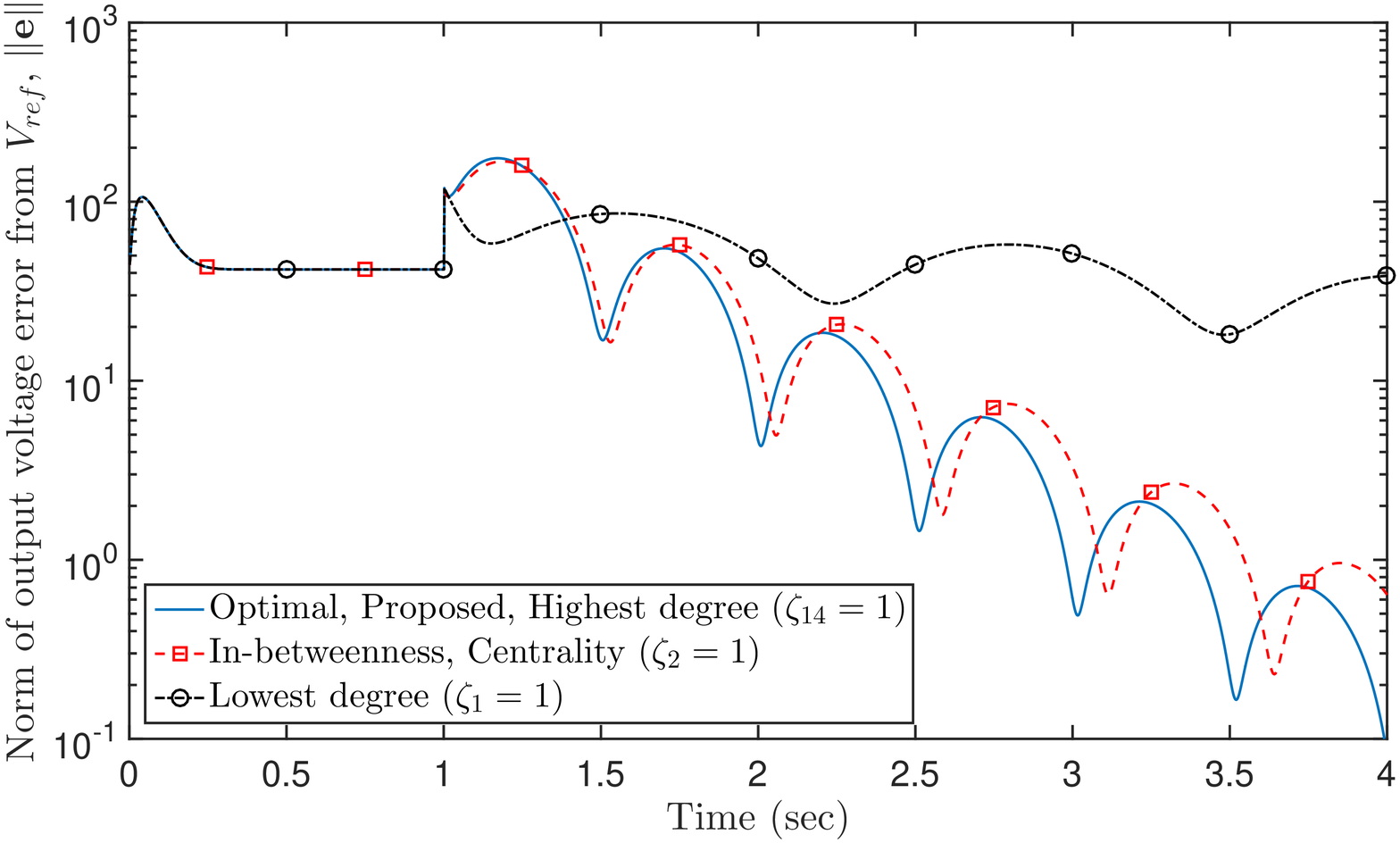} 
\caption{Evolution of norm of network error, $\Vert \text{\textbf{e}}\Vert $
, for several single pinning methods.}
\label{fig: SET}\vspace{-.5cm}
\end{figure}
Here, we consider the scenarios with single pinning. Table I gives the
results for solution of the problem in \eqref{eq: Problem1} for several
cases of node $i$ being pinned. In calculations of the second and fourth
columns, Theorem \ref{theorem: bounds_single_pinning} is used. The first row
of the table corresponds to the methods: optimal pinning, proposed algorithm
and high degree pinning methods, the second row shows the results for
highest in betweenness coefficient\footnote{%
Please see \cite{Klein10} for definitions.}, while the lowest degree pinning
method is given in the last row. Table I is sorted by descending $\mu _{N}$.
As can be observed, although the lower bounds, $\mu _{l}$, are trivial for
the case of single pinning for this particular example, the upper bounds, $%
\mu _{u}$, are very close to the actual value of algebraic connectivity, $%
\mu $. Furthermore, the proposed algorithm for this example yields the
optimal solution for the problem in \eqref{eq: Problem1}.

Fig. \ref{fig: SVT} shows the voltage signals for the synchronization
problem in islanded microgrid when DG 14 is pinned. Here, it is assumed that
at $t=0$[s], the microgrid is severed from the main grid and at $t=1$[s],
the secondary voltage control is applied. As can be seen, after the
microgrid is separated from the main grid, the output voltage of the DGs
synchronizes around a lower value than the grid value; this is due to the
distribution of the loads. However, after the secondary voltage control goes
online, all the output voltages converge to desired value of $v_{%
{\scriptsize \text{ref}}}=380$[Vrms].

In Fig. \ref{fig: SET}, the evolutions of norm of network error vector
defined in \eqref{eq: DCM} have been shown for various cases of single
pinning listed in Table \ref{table: single pinning}. It can be observed that
the lowest degree pinning results in the poorest convergence rate of output
voltage\textit{\ i.e.,} $\lambda =0.3$. The pinning based on highest
in-betweenness and centrality coefficients achieve convergence rates of $%
\lambda =1.9$. The optimal pinning, which also coincides with the solution
of our suboptimal algorithm, gives a convergence rate of $\lambda =2.2$%
\footnote{%
Since the plots are on a logarithmic (base 10)\ scale, the relationship $%
y=-\lambda t\log _{10}e$ holds for the exponentially convergent envelopes.
Thus, $\lambda $ can be inferred from a plot by multiplying the slope of its
envelope by a factor of $1/\log _{10}e\approx 2.3$, \textit{e.g.,} for the
optimal pinning plot in Fig. 3: $\lambda \approx \left( \log
_{10}(10^{2}/10^{-1})/(4-1)\right) \cdot \,2.3=2.3$ which is very close to
the reported accurate value of $\lambda =2.2$.}. 
\begin{figure}[t]
\includegraphics[scale = 0.255]{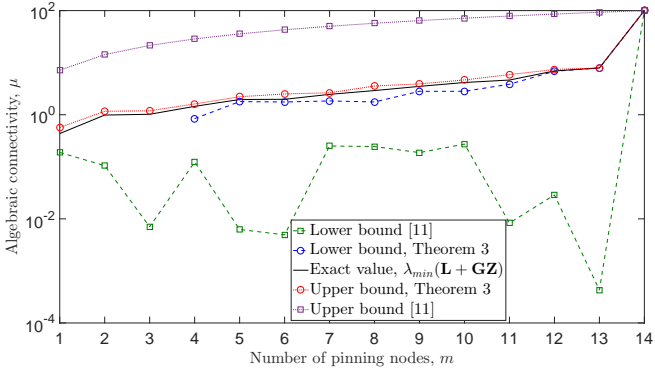}\caption{Lower and upper bounds for algebraic connectivity in case of
	optimal pinning: $g=100$.}
\label{fig: bounds optimal} \vspace{%
-.5cm}
\end{figure}

\subsection{Multiple Pinning}

For the scenarios of multiple pinning, we assume that $m=7$ nodes are
pinned. Table \ref{table: multiple pinning} gives the lower and upper bounds
as well as the value of objective function in (\ref{eq: objective}) for
several known pinning algorithms. The results in the first and second rows
correspond to optimal pinning selection and our proposed algorithm,
respectively. The third and fourth rows give the results for the lowest and
highest degree pinning methods while the fifth and sixth rows correspond to
highest centrality and highest in-betweenness pinning methods, respectively.
Table II is sorted by descending $\mu _{N}$. It can be observed that the
proposed algorithm outperforms the most common pinning methods.

\begin{figure}[t]
\includegraphics[scale = 0.26]{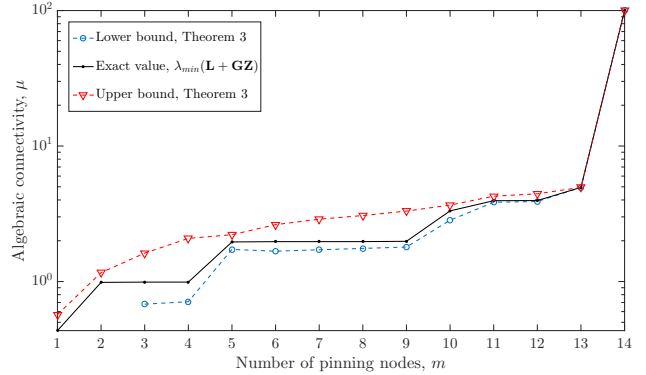}\caption{Lower and upper bounds for algebraic connectivity for the proposed
	algorithm.}
\label{fig: bounds suboptimal} \vspace{%
-.55cm}
\end{figure}

Fig. \ref{fig: bounds optimal} shows the two sets of lower and upper bounds
for optimal pinning as a function of the number of pinning nodes $m$. The
top and bottom plots (dashed-circle and dotted-line-square) are the upper
and lower bounds given in \cite{Pirani16}, respectively. The second from top
plot corresponds to the upper bound in (\ref{eq: upper_multiple})
(solid-line-solid circles) and the second from bottom plot
(solid-line-diamond) is the lower bound given in (\ref{eq: lower_multiple}).
It should be noted that the lower bound in (\ref{eq: lower_multiple})
results in $0$ for $m~=~1,~2,~3$. As can be seen, our lower and upper bounds
are close to the analytical value of algebraic connectivity and almost
always much better than those of \cite{Pirani16}. Also, it should be noted
that the lower bound given in \cite{Pirani16} requires the calculation of
the nonnegative eigenvector corresponding to the minimum eigenvalue, $%
\lambda _{\min }(\text{\textbf{L}}+\text{\textbf{G}}\text{\textbf{Z}})$.
Fig. \ref{fig: bounds suboptimal} gives the lower and upper bounds for our
algorithm as a function of the number of pinning nodes $m$. As it can be
observed, the lower and upper bounds are close to the analytical value of
algebraic connectivity.

\begin{figure}[tbp]
\includegraphics[scale = 0.26]{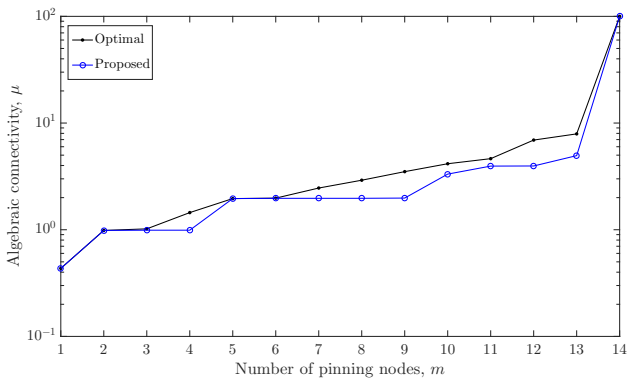} \caption{Algebraic connectivity to pinning signal for optimal pinning and
	the proposed algorithms.}
\label{fig: optimal vs suboptimal}
\vspace{-.5cm}
\end{figure}

Fig. \ref{fig: optimal vs suboptimal} compares the proposed method to the
optimal pinning method. It can be seen that the proposed algorithm closely
follows the optimal pinning method. It should be also noted that if $m=7$,
the number of searches to find the optimal solution is $3432$ compared to $%
77 $ for the proposed algorithm. This shows how our algorithm drastically
reduces the complexity of the solution for the pinning problem. 

\begin{figure}[t]
\includegraphics[scale = 0.265]{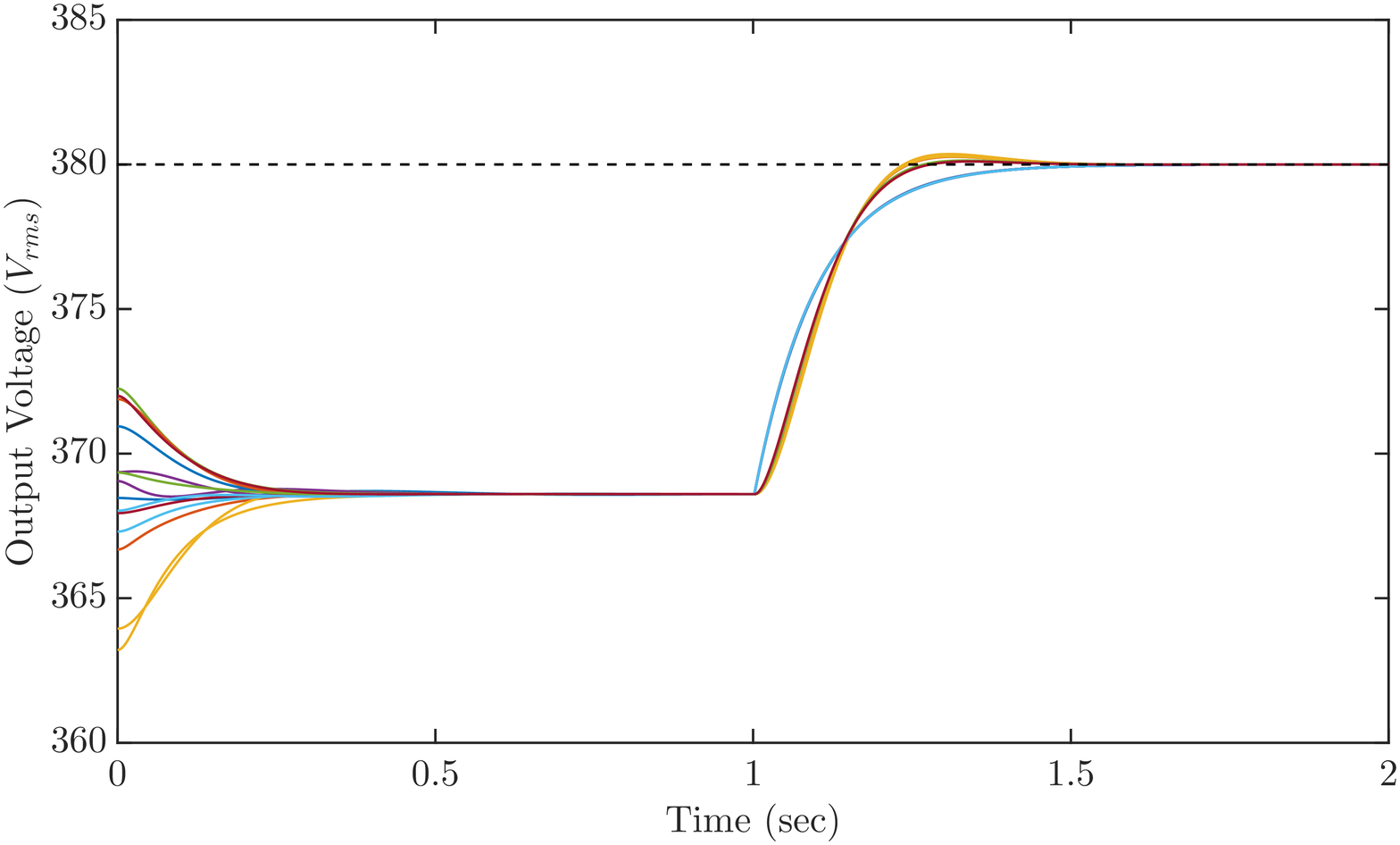}\caption{Evolution of voltage signals of the microgrid under optimal
	pinning, $m_{0}=7$.\newline
}
\label{fig: MVT}%
\includegraphics[scale
= 0.276]{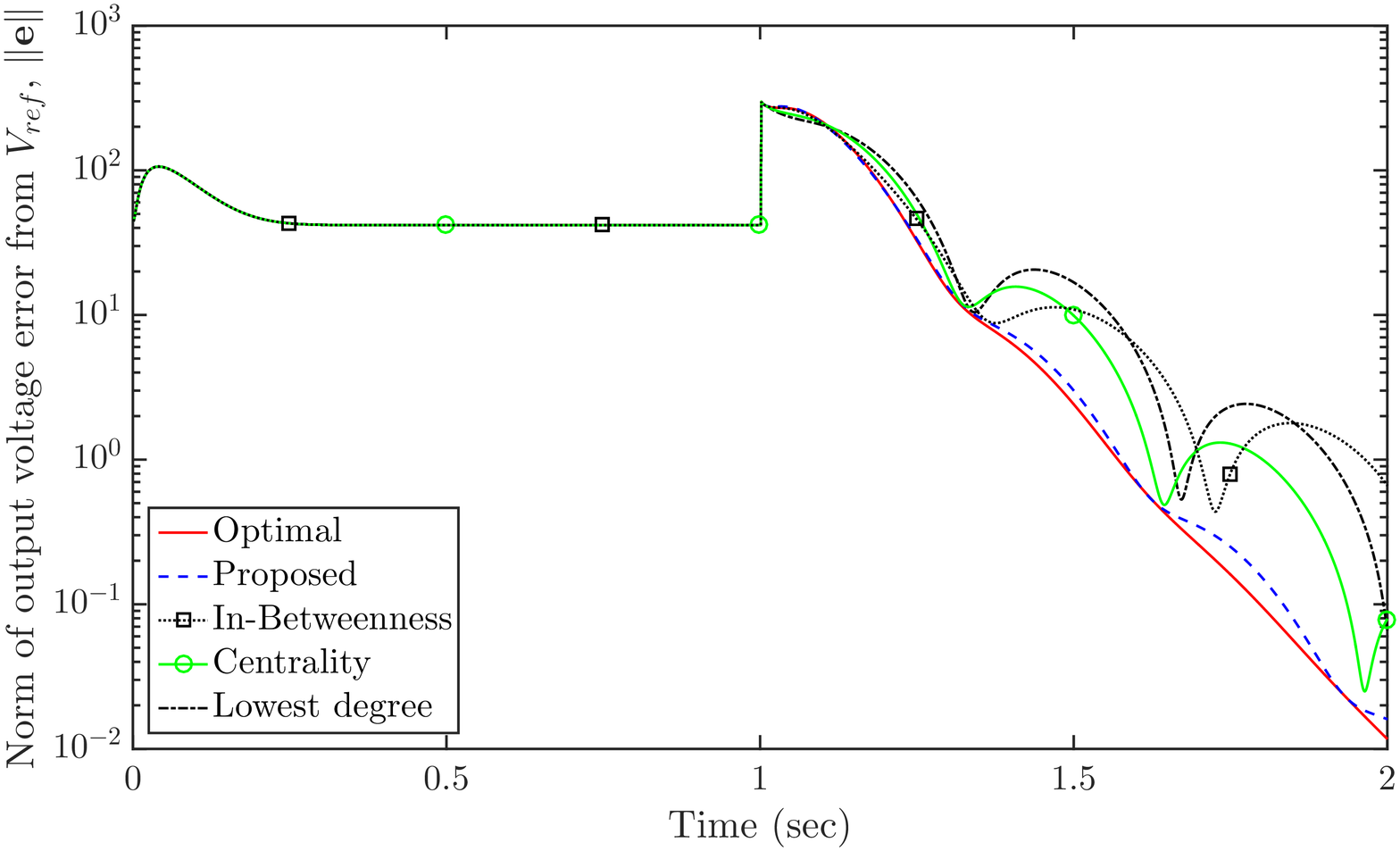}
\caption{Evolution of norm of network error, $\Vert \text{\textbf{e}}\Vert $
, for several pinning methods, with $m_{0}=7$, listed in Table \protect\ref%
{table: multiple pinning}. }
\label{fig: MET} \vspace{-.3cm}
\end{figure}

Similar to Fig. \ref{fig: SVT}, in Fig. \ref{fig: MVT}, we have assumed that
at $t=0$[s]$,$ the microgrid is severed from the main grid and at $t=1$[s]$,$
the secondary voltage control is applied. Aside from observations similar to
those made for Fig. \ref{fig: SVT} earlier, it is clear that increasing the
number of pinning nodes results in improved transient behavior. Fig. \ref%
{fig: MET} provides more quantitative results on the convergence rate for
different pinning methods when the number of pinned nodes is $m=7$. By
calculating the rate of exponential decay for the envelope of the norm of
voltage error-vector defined in \eqref{eq: DCM}, the convergence rates for
high in-betweenness and centrality methods are $\lambda =4.95$ and $\lambda
=4.97$, respectively. The convergence rate for lowest degree algorithm is $%
\lambda =6.50$. Finally, the convergence rate of the proposed algorithm can
be obtained as $\lambda =9.95$, whereas for the optimal pinning, $\lambda
=10.21$ is achieved. Considering the amount of the reduction in complexity
from the optimal algorithm to the proposed algorithm, the small loss in
performance, \textit{i.e.,} convergence rate, can be justified for most
applications. It should be noted that although the errors for all cases are
very small after a few seconds, for a power grid application, if the errors
do not settle between $-5\%$ to $+10\%$ of the reference voltage within 10
to 20 cycles ($0.15-0.30$ seconds for $f=60Hz$), the protective relays will
operate and remove the DG(s) from the grid.

\section{Conclusions}

In this paper, we first derived analytical lower and upper bounds on the
algebraic connectivity of pinned networked systems. Analyzing these bounds,
several limitations of pinning control on algebraic connectivity with
respect to reference state were shown. Next, based on the bounds, we formed
an objective function to propose a suboptimal algorithm with polynomial
complexity. Numerical examples have shown that the derived upper and lower
bounds closely track the value of algebraic connectivity as the number of
pinning nodes is varied from single pinning through to an all-nodes-pinned
system. Finally, the application of the proposed algorithm to design the
secondary voltage synchronization control in a network of distributed
generators in a microgrid operating in islanded mode illustrates its
efficacy for both single and multiple pinning scenarios.

\bibliographystyle{IEEETran}
\bibliography{mybib}

\end{document}